\documentclass[english]{article}
\usepackage[T1]{fontenc}
\usepackage[latin9]{inputenc}
\usepackage{geometry}
\usepackage{color}
\usepackage{url}
\usepackage{graphicx}

\makeatletter

\providecommand{\tabularnewline}{\\}

\makeatother

\usepackage{babel}
\begin{document}
\title{Networks of Twin Peaks: the Dale Cooper Effect}
\author{Harun \v Siljak, Trinity College Dublin, Ireland}
\date{}
\maketitle

\section{The shape of stories}

\begin{figure}
\begin{centering}
\includegraphics[width=0.9\textwidth]{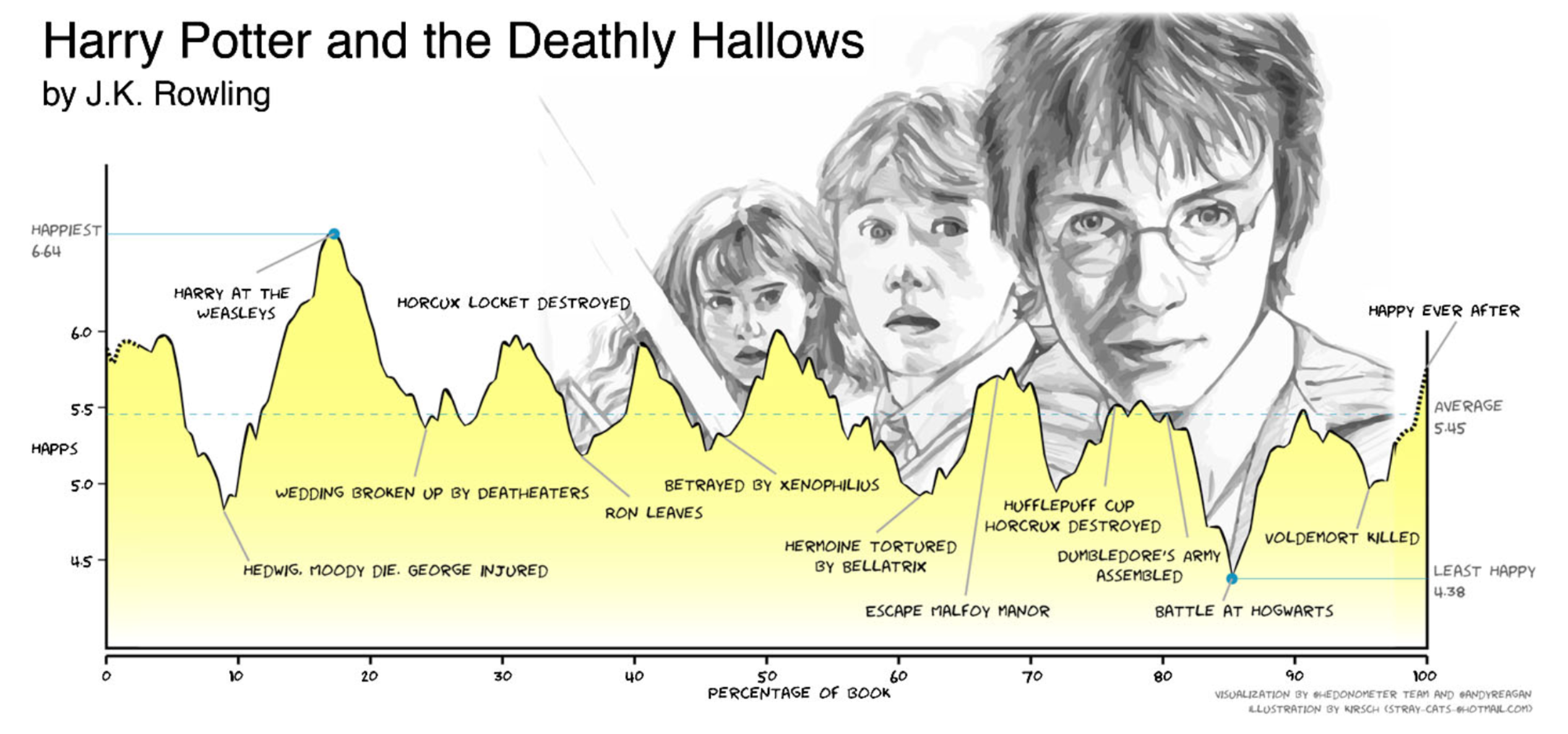}
\par\end{centering}
\caption{Annotated emotional arc ofHarry Potter and the Deathly Hallows, by
JK Rowling (from \cite{key-2})}
\label{f1-2}
\end{figure}

When Kurt Vonnegut speaks of the shape of stories \cite{key-1}, he
is in the realm of calculus, continuous functions describing dynamics
of the plot, the highs and lows of the protagonist's emotions, the
speed at which the story unravels. In a way, it corresponds to the
historical development of calculus as a way to interpret motion as
a function of time. The abscissa is the time, and the ordinate axis
is something important for the hero--emotional state, for example.
Figure \ref{f1-2} provides a detailed illustration from contemporary
fantasy fiction.

There is another shape of a story we are interested in: one of its
people, the world it lives in. That shape does not focus on the actions,
but on the structures. It describes the contents of the carrier bag,
like in Ursula K. Le Guin's Carrier Bag Theory of Fiction \cite{key-3}\textcolor{black}{,
and they are neatly arranged in graphs, maps, and trees (as the eponymous
work by Franco Moretti suggests \cite{key-4}).} The tools from it
come from graph theory and algebra--and once put in context of arbitrary
large networks, the field of research is usually dubbed network theory
\cite{key-5}. Are there only so many community types, networks of
interactions in life and in literature? What can mathematics tell
us about it? Are networks of characters in fiction anything like our
social networks? A growing body of literature has been dealing with
this application of network theory; investigating epic stories from
the past, superhero universes from the modern era, etc. The more characters
exist in a work of fiction, the more interesting and relevant network
theoretic results become: patterns emerge, clusters and communities,
and unexpected characters reveal themselves as keystone elements of
networks: once removed, everything falls apart.

\textcolor{black}{``I would go so far as to say that the natural,
proper, fitting shape of the novel might be that of a sack, a bag.
A book holds words. Words hold things. They bear meanings. A novel
is a medicine bundle, holding things in a particular, powerful relation
to one another and to us. '' \cite{key-4} Here, in this quote from
Le Guin lies our motivation to explore the networks, rather than forward-driving,
sharp-edged plots of Vonnegut. She proceeds: ``{[}I{]}t\textquoteright s
clear that the Hero does not look well in this bag. (...) That is
why I like novels: instead of heroes they have people in them.''
We are curious what the hero effect is in a network: are they just
like anyone else, do they fit in the bag? Finally, we erase the linear
time of Vonnegut's plots by observing networks instead of actions,
algebra instead of calculus. The community in the network exists independently
of the passage of time, and it needs to feel real. Again, Le Guin
writes ``If, however, one avoids the linear, progressive, Time\textquoteright s-(killing)-arrow
(...) one pleasant side effect is that science fiction can be seen
as (...) in fact less a mythological genre than a realistic one.''}

In this work, we scratch the surface of character networks in the
cult TV show \emph{Twin Peaks}. For the readers who have not seen
it, the first season of Twin Peaks revolves around a murder investigation
in the town of Twin Peaks. FBI agent Dale Cooper comes to town to
investigate alongside the local police department, and the viewers
get to see networks of crime, romantic relationships, and law enforcement
entangle and blur the lines between. Compared to other network theoretic
treatises, this network is small scale, with sixty characters introduced
over the course of eight episodes. The community is not even expected
to behave as a social network; the crime-investigation-driven plot
of \emph{Twin Peaks }was expected to merely shine the light on those
relationships relevant to the murder of Laura Palmer, hyperfocusing
on a network built by the law enforcement officers conducting the
investigation. Not only did the network analysis confirm that the
community is more than just a set of photos pinned to a plywood board
at the Sheriff's office: we also discovered a new storytelling network
phenomenon we called the Dale Cooper Effect, a phase transition in
network structure.

\section{Graphs and associated tools}

\begin{figure}
\begin{centering}
\includegraphics[width=0.7\textwidth]{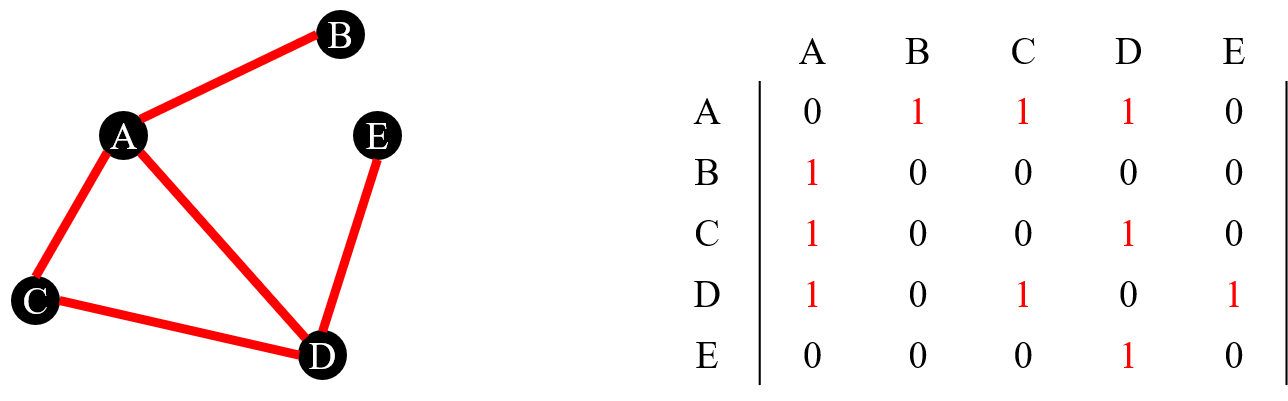}
\par\end{centering}
\caption{A graph and its adjacency matrix}
\label{f1}
\end{figure}

A graph is a mathematical object made of vertices and edges (Fig.
\ref{f1} left), which in the context of networks are often called
nodes and links, respectively. A pair of vertices can be connected
by a single edge or no edges--e.g. B and E have no edge connecting
them, while A and B do. For large graphs and for algebraic manipulation
of them, a commonly used representation is that of the adjacency matrix
(Fig. \ref{f1} right). It is a square matrix whose dimension corresponds
to the number of vertices in the graph. As Fig. \ref{f1} shows, if
there is an edge between the vertex corresponding to the row, and
the one corresponding to the column, we place a ``1'' at their intersection
in the matrix. Otherwise, that element is a ``0'' (no edge). If
the graphs have no loops (i.e. if connecting vertex A to itself via
an edge is not an option), the diagonal elements of the adjacency
matrix are all zero. Furthermore, if the graph is undirected (i.e.
if an edge between A and B is interpreted both as a connection of
A with B and B with A), the matrix is symmetric.

\begin{figure}
\begin{centering}
\includegraphics[width=0.7\textwidth]{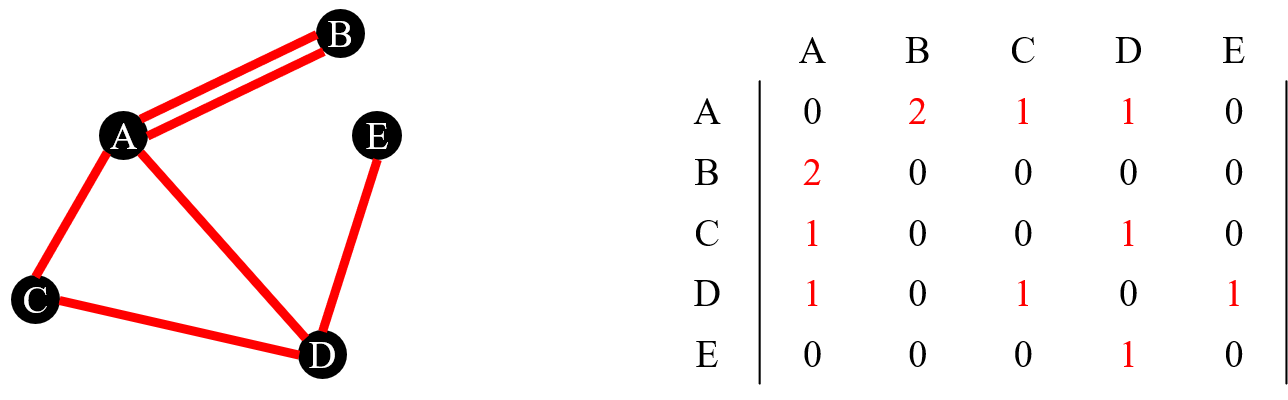}
\par\end{centering}
\caption{A multigraph and its adjacency matrix}
\label{f1-1}
\end{figure}

An extension of the concept of the graph is a multigraph, where multiple
edges connecting same two vertices are allowed. An example of this
is shown in Fig. \ref{f1-1}, where vertices A and B now have two
common edges. This maps directly into the adjacency matrix, and the
entries AB and BA are now equal to two. This multigraph can also be
interpreted in terms of weighted graphs: all edges would again be
single, but the one between A and B would have the weight of 2, while
others would have the weight of 1. Sometimes, for example in collaboration
networks where the goal is to assess the distance between entities
represented by nodes (e.g. co-authors in mathematics or co-stars in
movies, c.f. Erd\H{o}s number, Bacon number), $n$ multiple links
between two nodes can be taken to correspond to the weight of $1/n$.

At this point, it is useful to define the degree of a vertex in a
graph: for undirected graphs it is simply the number of edges connected
with it (i.e. the sum of the corresponding row/column in the adjacency
matrix). In the example in Fig. \ref{f1}, the degree of nodes A and
D is 3, the degree of the node C is 2, and the degree of nodes B and
E is 1.

When drawing conclusions from networks appearing in the world around
us, it is often useful to compare their structure to that of basic
network models; common features stand out and cross over in different
contexts, suggesting general patterns. Two common random network models
are the Erd\H{o}s-Renyi (random network) and the Barabasi-Albert (preferential
attachment) model.

In the Erd\H{o}s-Renyi model, the adjacency matrix is filled by tossing
a biased coin. Let it be a $200\times200$ matrix, and let the probability
of two vertices being connected by an edge be $0.1$. The upper triangle
of the adjacency matrix is then populated entry by entry with the
result of tossing a coin which has 10\% chance of heads (1), and 90\%
tails (0). After the upper triangle is filled, it is symmetrically
copied into the lower triangle, finalising the undirected Erd\H{o}s-Renyi
graph generation. 

The Barabasi-Albert model recognises the observation that many networks
in the real world are built with the mechanism of preferential attachment.
When a new node is added to such a network, the links it forms are
not drawn with equal probability from the set of all nodes already
in the network. The nodes that are already well-connected with other
nodes will have more chance in attracting the new node. Let our network
once again have 200 nodes; in the process of preferential attachment
we will be adding one node at a time, assign 4 links connecting it
to existing nodes in the network, and end the process once all 200
nodes are added. The probability that the newly added node will establish
a link with a particular existing node is proportional to the existing
node's degree.

\begin{figure}
\begin{centering}
\includegraphics[width=0.7\textwidth]{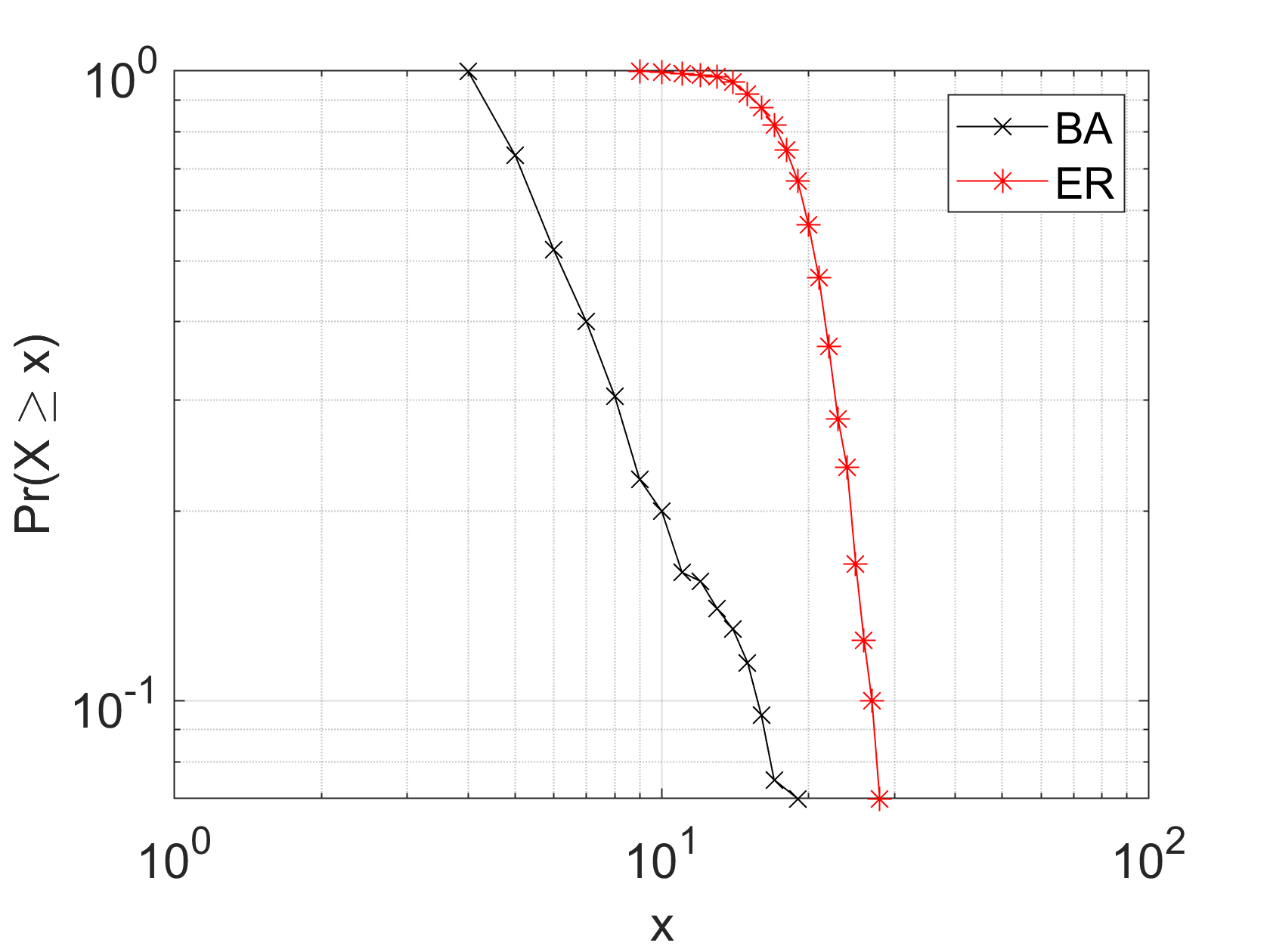}
\par\end{centering}
\caption{Examples of degree distributions for random networks}
\label{f1-1-2}
\end{figure}

In Figure \ref{f1-1-2} we depict the complementary cumulative distribution
function of the degrees in two network model examples we described.
The plot represents the probability that a node in the network has
degree $X$ greater than some value $x$. For an $n$-node random
network, the degree of a node is a function of $n-1$ coin tosses
and hence is a function of scale: 50\% of the nodes will be expected
to have, in our numerical example, degree lower than $200\cdot0.1=20$,
and 50\% will be expected to be over it. However, for the preferential
attachment network, no such scale exists as the network generative
process could have continued beyond $n$ nodes (this is why we often
refer to these networks as scale-free). The complementary cumulative
distribution function for such a network is theoretically a straight
line in a log-log scale plot (here we present the actual plot from
200-node network to show the deviations seen at small scales for a
single instance of a network). \textcolor{black}{While the ``preferential
attachment results in power law'' maxime often holds, interesting
exceptions have been observed--a famous one is the citation distribution
of Physical Review papers demonstrated by Sid Redner to follow a log-normal
distribution \cite{key-6}.}

Another measure of interest for us in this note is that of assortativity.
Are nodes with large/small degrees more likely to be connected to
other vertices with large/small degrees in a network? In real-world
social network, it is expected to see more popular nodes of the network
to cluster together, while in networks such as the internet, webpages
with few links to other webpages are more likely to be connected to
well-linked hubs. The former is the case of a positive assortativity
coefficient, and the latter of a negative one. Random networks, such
as ones created by processes introduced earlier in this section, are
expected to have a neutral (zero) assortativity.

The common reason to investigate networks in works of fiction is to
compare their statistical properties to those of real networks. For
example, in \cite{key-7} the authors apply the network theory tools
to study networks of characters in the epic narratives of The Beowulf,
The T\'ain B\'o C\'uailnge and The Iliad. The network of the Iliad was found
to have features of realistic networks such as positive assortativity,
scale-free behaviour. For Beowulf, these properties are achieved once
the protagonist, Beowulf, is removed from the story--Beowulf's disproportionate
degree of interaction with the world otherwise skews it. For The T\'ain
B\'o C\'uailnge, the key to reaching realistic network properties has
been removing one-off interactions between the six protagonists of
the epic and the remainder of the world. There is always a component
of ``historicity evaluation'' when discussing the studies of epics--assessing
the realness of social networks in them is a factor in thinking about
their genesis. For \emph{Twin Peaks, }we have no such concerns; it
is a fantasy world for which we do not expect to be a snapshot of
the society. However, we are curious to see what the storytelling
structure reveals.

\section{Twin Peaks}

To investigate the networks of interaction in \emph{Twin Peaks, }we
analysed the transcripts from season 1 of the show (total of 8 episodes)\cite{key-8}.
The choice to restrict the dataset to season 1 was motivated by the
evolution of the script and cast in season 2, followed by the constraints
passage of time put on season 3 (filmed two decades after the first
two). Season 1 was hypothesised to be the smallest homogeneous unit
appropriate for analysis. The adjacency matrix was formed chronologically:
every time a new character appears, a new node is added to the graph.
For every scene two characters share screentime in, a new edge is
added to the graph, allowing multiple edges between the same pair
of nodes. If three characters appear in the scene, all pairs get a
new edge.

\begin{figure}
\begin{centering}
\includegraphics[width=0.45\textwidth]{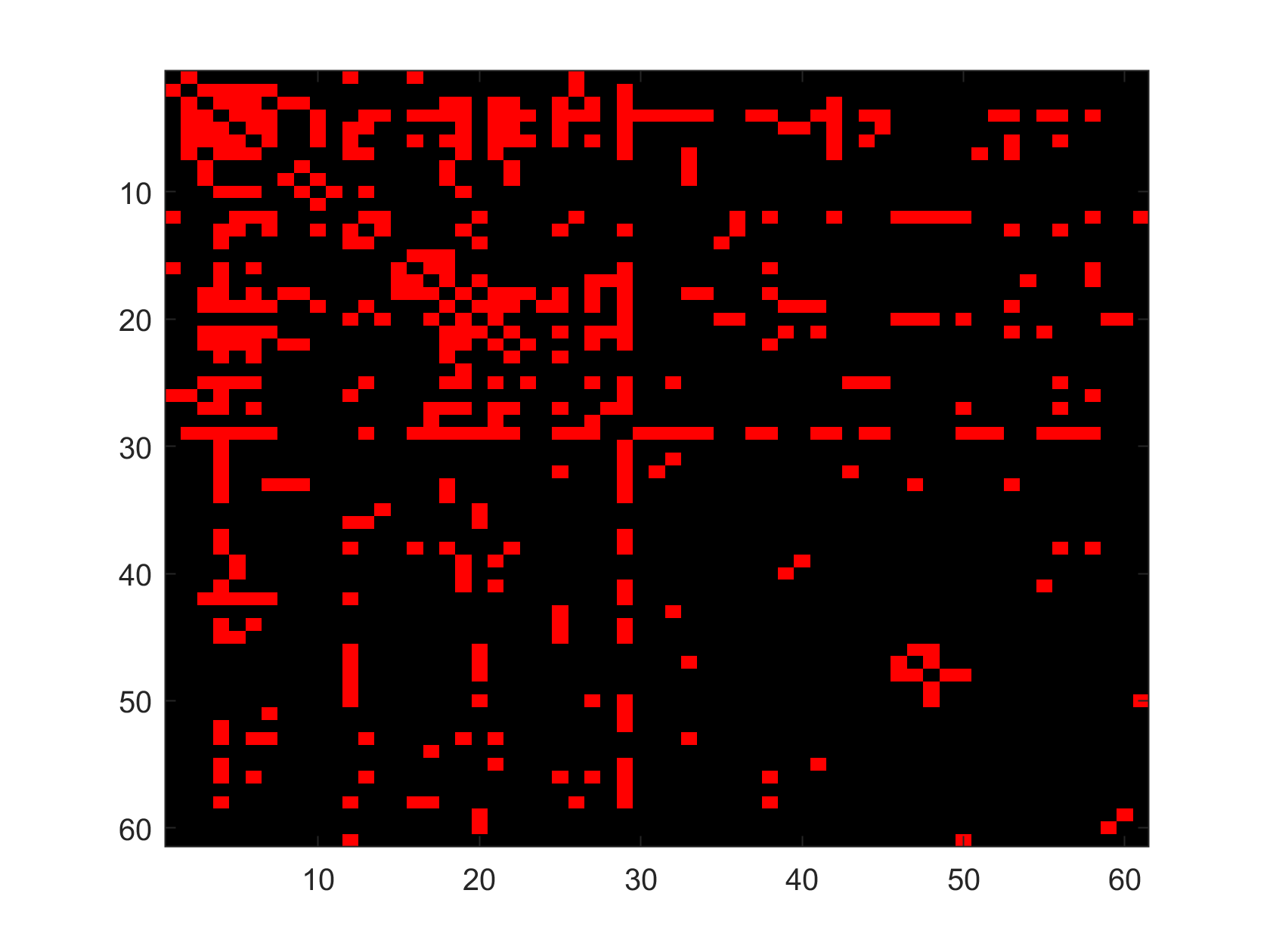}\includegraphics[width=0.45\textwidth]{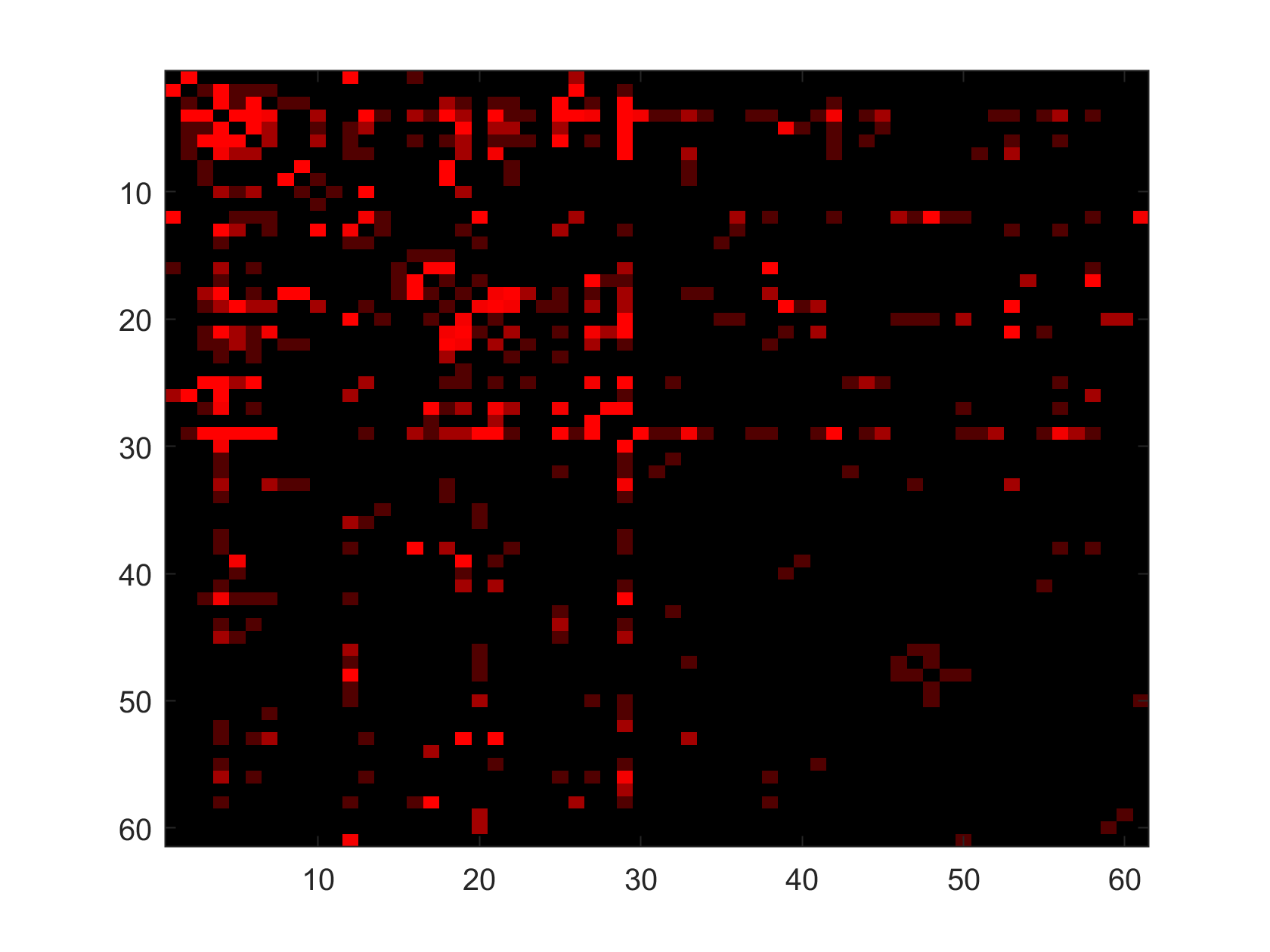}
\par\end{centering}
\caption{Adjacency matrices from Season 1 of Twin Peaks: graph (left) and multigraph
(right)}
\label{f1-1-1}
\end{figure}

The resulting matrix is graphically presented in Fig. \ref{f1-1-1}:\textcolor{blue}{{}
}\textcolor{black}{our visualisation is similar to that of the Les
Miserables Co-occurrence \cite{key-9}--the difference is in ordering,
as we preserve the order of appearance in the matrix.} The left representation
corresponds to the binary interpretation of the graph seen in Fig.
\ref{f1}: characters either share screentime, or they don't, the
quantity of time is irrelevant. Red squares correspond to edges of
the graph, i.e. characters that are connected. On the right, the shade
of red changes with respect to the number of scenes characters have
in common (i.e. interpretation akin to Fig. \ref{f1-1})--brighter
means more interactions.

\begin{figure}
\begin{centering}
\includegraphics[width=10cm]{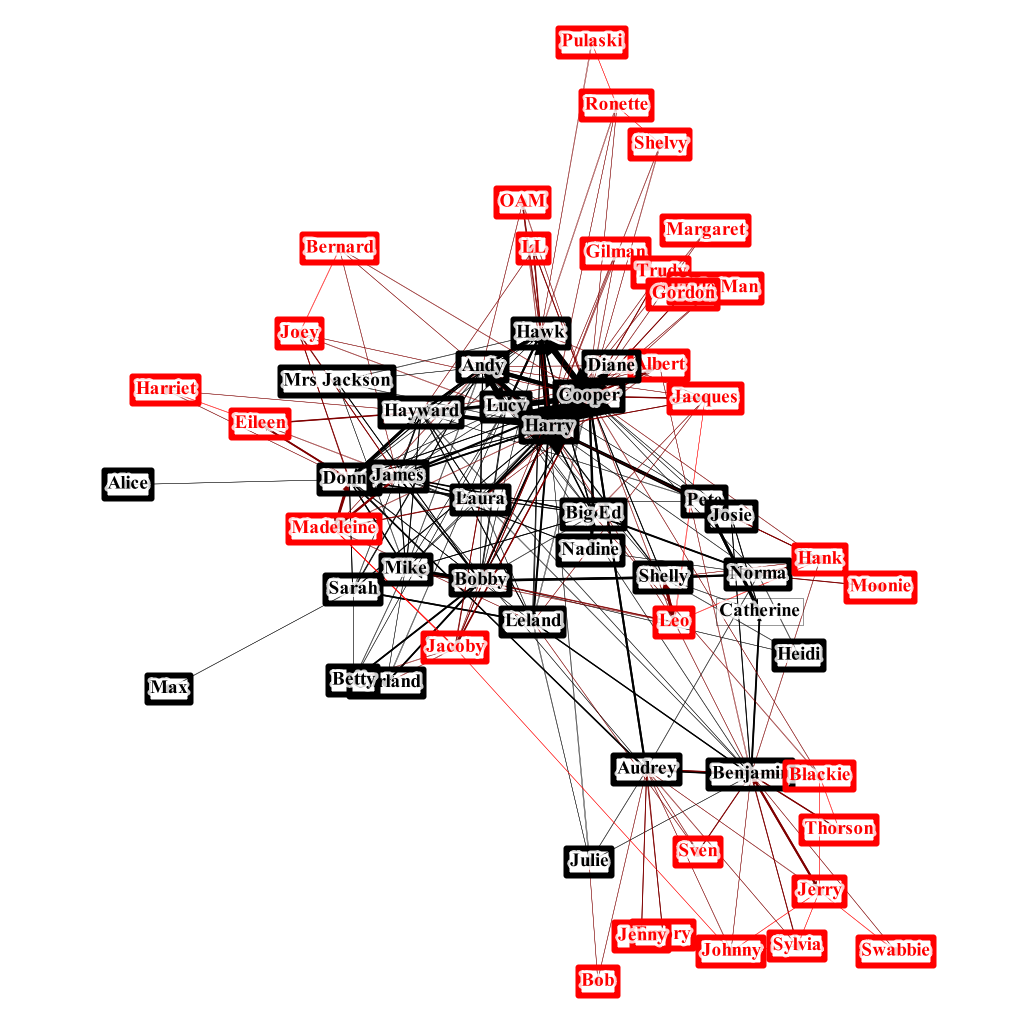}
\par\end{centering}
\caption{The graphical representation of the Twin Peaks network:\textcolor{blue}{{}
}\textcolor{black}{the black nodes are BC, the red are AD}}
\label{f1-1-1-1-1-1}
\end{figure}

We recognise that the characteristic ``cross'' in the middle of
the matrix, dividing it into four quadrants is the protagonist of
the show, \emph{Special Agent Dale Cooper}. Cooper appears 36 minutes
into the pilot (first episode of the show), and he is the median character:
$\approx30$ characters are introduced before him, and $\approx30$
after him (as we stated already, the adjacency matrix is filled chronologically,
so $n$th row/column of it correspond to the $n$th character to appear
in the show). If we now divide the characters into two categories--those
appearing before Cooper (denoted BC in the remainder of this paper)
and those after Dale (AD). The brightest points in the right adjacency
matrix are usually either couples, or the Twin Peaks Police Department
combined with Cooper. Graphically, this is shown in Fig. \ref{f1-1-1-1-1-1},
while the most popular nodes and links are tabulated in Table \ref{tabela}.

\begin{table}
\begin{centering}
\begin{tabular}{|c|c|c|c|c|}
\hline 
 & The most connected characters & \# & The most repeated links & \#\tabularnewline
\hline 
\hline 
1 & Dale Cooper & 35 & Cooper \& Truman & 47\tabularnewline
\hline 
2 & Sheriff Truman & 35 & Truman \& Lucy & 16\tabularnewline
\hline 
3 & Deputy Andy & 20 & Cooper \& Hawk & 15\tabularnewline
\hline 
4 & Donna & 19 & Truman \& Hawk & 14\tabularnewline
\hline 
5 & Bobby & 18 & Donna \& James & 13\tabularnewline
\hline 
\end{tabular}\label{tabela}
\par\end{centering}
\caption{Popular links and nodes in the network: all of them are in the BC
part. The most connected AD character appears as the 18th in the list;
the top repeated link which includes an AD character appears as the
13th in the list.}
\end{table}

\begin{figure}
\begin{centering}
\includegraphics[width=7cm]{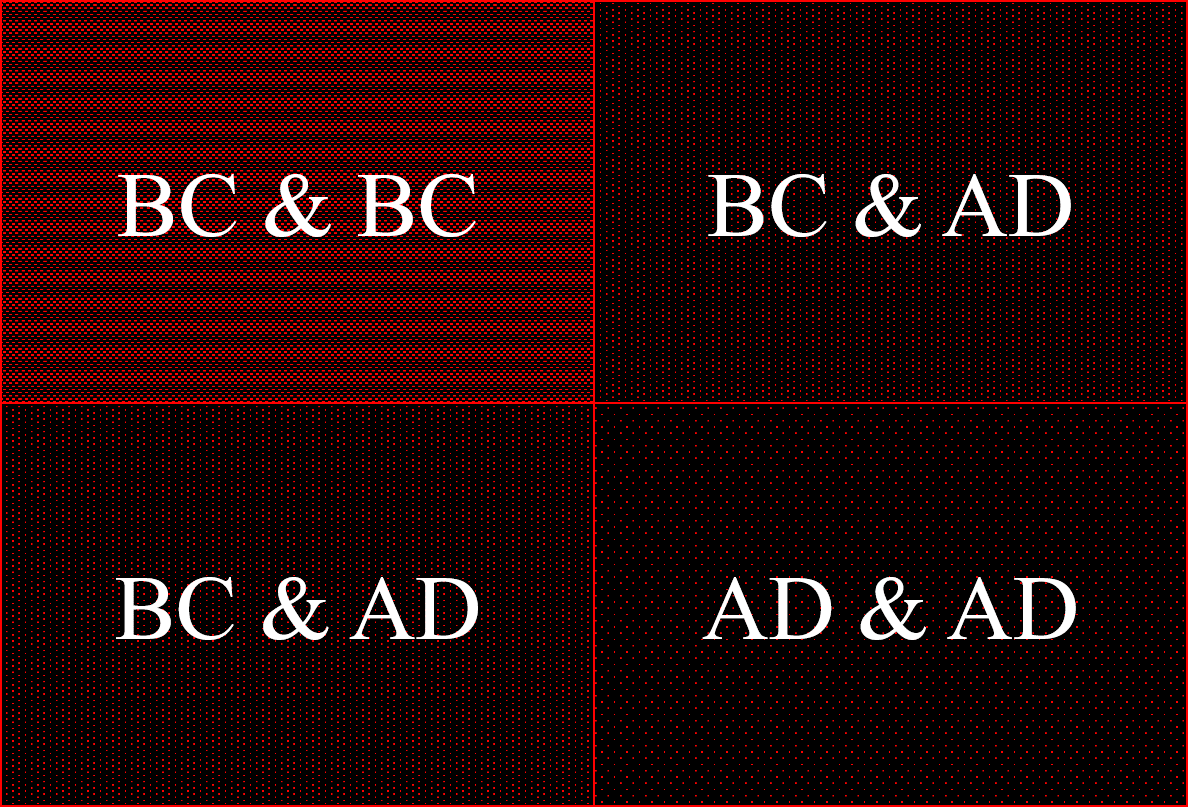}
\par\end{centering}
\caption{The Dale Cooper Effect}
\label{f1-1-1-1-1}
\end{figure}

In Fig. \ref{f1-1-1-1-1} we discuss the relationship between networks
of BC characters and AD characters, what we call \emph{The Dale Cooper
Effect}. This cartoon version of the adjacency matrix aims to show
how BC is a closely knit community--the part of the adjacency matrix
corresponding to it (top left quadrant) is densely filled with connections.
On the other end, the AD characters interact among themselves very
little, the bottom right quadrant is very sparse in connections. It
is by virtue of interacting with BC characters that AD characters
are a part of the network (bottom left and top right quadrants).

This suggests a mechanism by which the two networks are spliced together.
After the nodes of BC network are set up, and Dale Cooper is introduced,
new AD nodes are added with a preferential attachment to BC: a newly
added node has a much higher probability of being connected to the
BC core than fellow AD periphery. Also, the density of connections
is indicative: the full graph has 236 edges, while the BC graph (in
this treatise, we include Cooper in BC) has 128. Doubling the number
of characters (i.e. adding AD to BC) hence results in doubling the
number of edges (linear scaling).

The mechanism of network growth in the previous paragraph inspires
a dive into the degree distribution of the network. Cooper and Sheriff
Truman appear together in many scenes and share most of their contacts:
their degree is the same, and rather high (35 in the entire network,
i.e. more connected with more than 1/2 of other characters; 20 in
the BC network, growing the fraction to 2/3 of other characters).
Motivated by the findings from Beowulf and the T\'ain where adjustments
are needed to make the protagonists more realistic, we were curious
to see if such high degree nodes as Cooper and Truman are expected
in networks like these.

\begin{figure}
\begin{centering}
\includegraphics[width=0.7\textwidth]{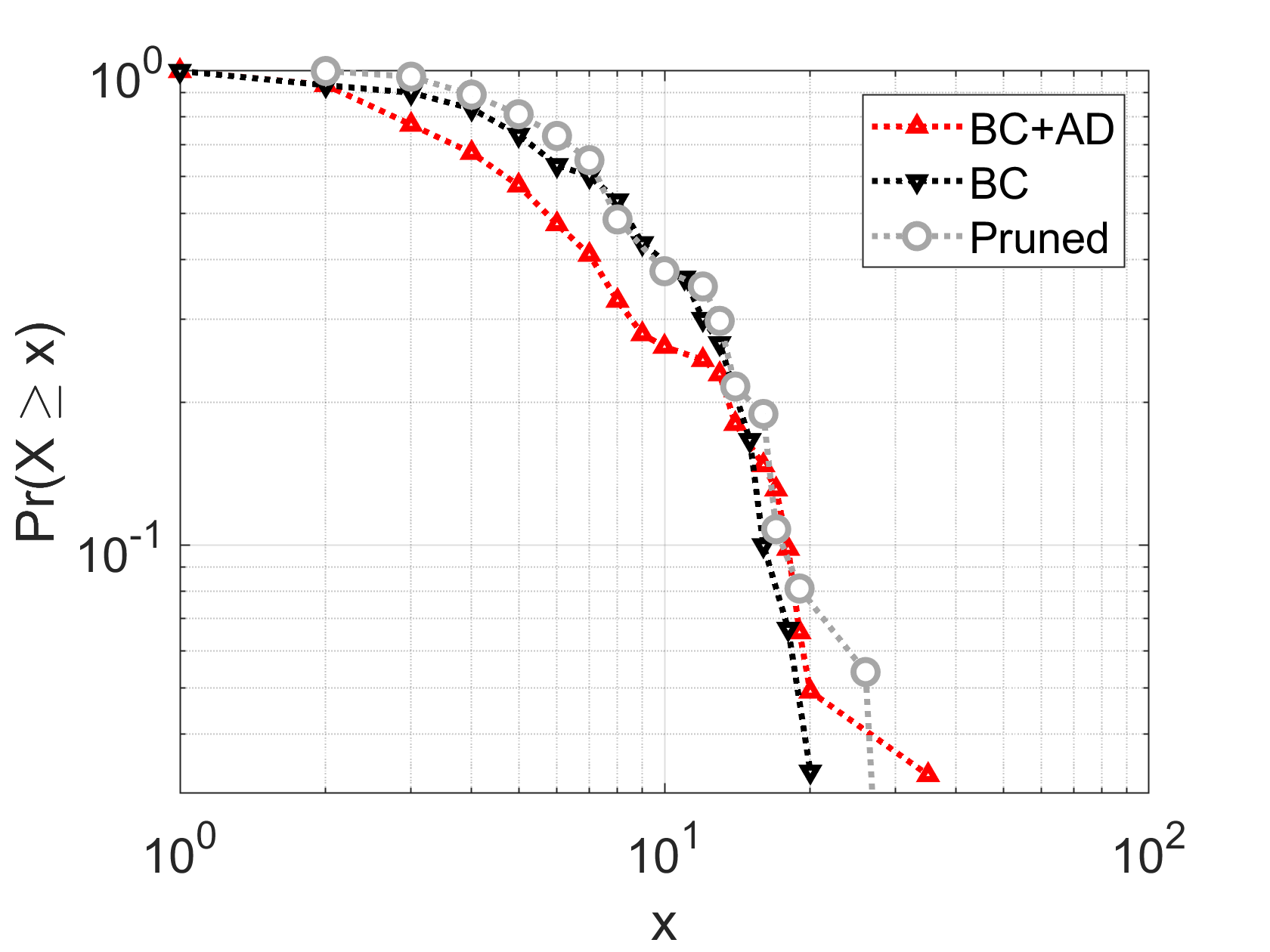}
\par\end{centering}
\caption{Degree distribution in the Twin Peaks networks compared to random
networks}
\label{f1-1-1-1}
\end{figure}

Figure \ref{f1-1-1-1} presents our findings. First, we note that
the BC degree distribution is similar to one of a random Erdos Renyi
network seen in \ref{f1-1-2},\textcolor{black}{{} and not too far off
from a log-normal distribution (whose surprising appearance in some
preferential attachment networks we already mentioned)}. For the BC+AD
network, we examine significant deviation from the shape: the curve
appears to be a superposition of three segments: one for low degrees
($x<10$), one for moderately high degrees ($x<20$) and one for very
high degrees ($x>30$). This is the effect of the mechanism by which
AD characters are connected: they dominate the low degree segment
of the distribution and shift it leftwards, while contributing to
an increase in degree for Cooper and Truman (the very high degrees
data point in the plot), pushing them rightwards. From here we deduce
that Truman and Cooper are not abnormally well connected in the BC
network, but in the BC+AD network, they are more popular than they
would be in a random network.

This makes sense as Truman and Cooper drive the plot in the AD introduction
period, and are often the first point of contact for new characters.
This is mirrored in the assortivity measures for BC and BC+AD networks:
BC assortativity is approximately zero (neutral), while for BC+AD
the assortativity is negative: less connected characters in AD are
connected to popular BC characters.

Following the ideas from \cite{key-7}, we proceed with studying the
options of character removal. The first question is about the effect
of removing characters that appear in a single scene (correction similar
to that applied to the \emph{T\'ain B\'o C\'uailnge} in \cite{key-7}).
This is the pruned network in Fig. \ref{f1-1-1-1}: the removed nodes
are predominantly in the AD group (20 removed from AD, and 4 from
BC), so it is not surprising that the pruned network gets closer to
the BC degree distribution. However, the anomaly of Cooper and Truman
remains. As it turns out, the pruned network collapses onto BC if
one of the two dominant nodes is removed: a case for ``this town
ain't big enough for both of us''. In a way, removing Cooper could
be justified as removing the outsider (or \emph{Beowulf}, following
on \cite{key-7}) from the local network of Twin Peaks.

Finally, given the importance of the law enforcement characters in
a show driven by crime investigation, an interesting test is whether
the BC+AD network remains connected after the members of the FBI and
the Twin Peaks Police Department are removed. The answer is yes--it
still remains a single graph (with the exception of the few characters
whose appearance in the show, if limited to Season 1, is tied exclusively
to the police investigation) with Benjamin Horne as its central figure,
tying together the young population of Twin Peaks through his daughter
Audrey, and the criminal milieu through his business connections.

\begin{figure}
\begin{centering}
\includegraphics[width=0.45\textwidth]{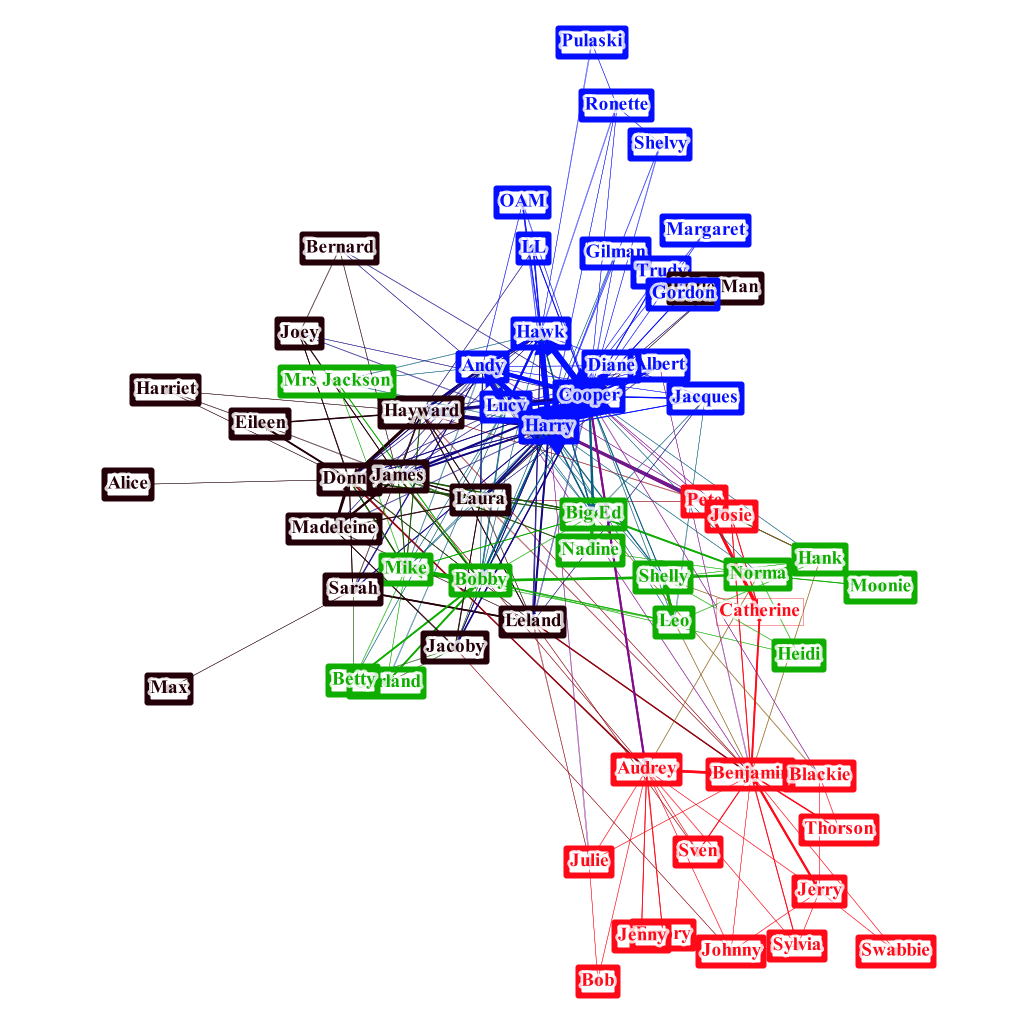}\includegraphics[width=0.45\textwidth]{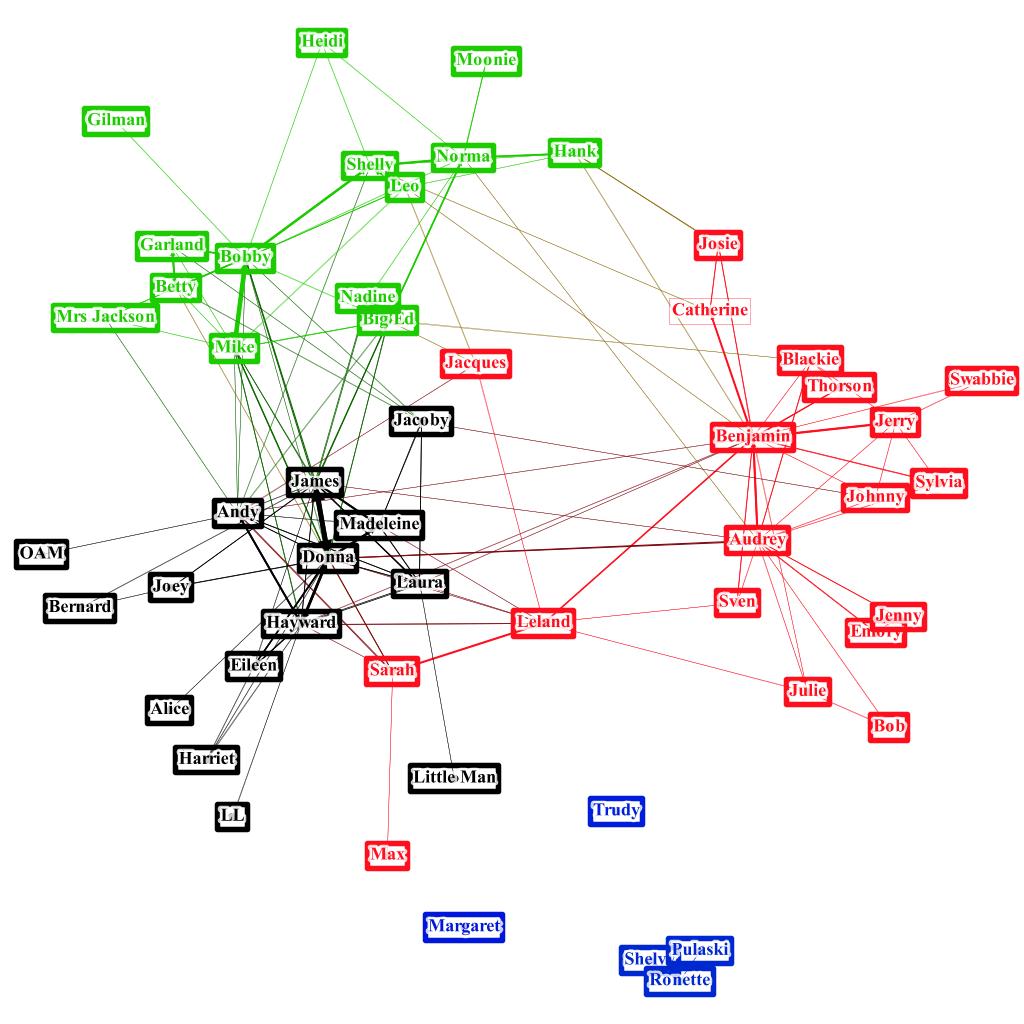}
\par\end{centering}
\caption{\textcolor{black}{Clusters in the Twin Peaks network before (left)
and after (right) removing the law enforcement}}
\label{f1-1-1-2}
\end{figure}

\textcolor{black}{This process of ``police removal'' is interesting
from the clustering point of view. Networks can often be split into
meaningful clusters of nodes based on the connections between them.
Sometimes this split is ambiguous and it is hard to deal with borderline
cases, but they give a general idea of the communities existing in
the network. Using the Louvain algorithm \cite{key-10} we perform
clustering of both the network before and after removal of police
officers in Twin Peaks, with the results shown in Fig. \ref{f1-1-1-2}.
The blue cluster on the left represents the cluster consisting of
the police and people primarily interacting with the police, either
as suspects or witnesses. Once the police officers are removed from
the network, some nodes are left disconnected, and the rest of the
nodes that had connections with other clusters join those clusters.}

\section{Final Remarks}

It is hard to draw conclusions from small data and somewhat arbitrary
conventions applied in its analysis. Given that we are in the realm
of art, creation, and interactions, one alternative to the approach
taken in this work would have been resorting to the actor network
theory (despite the fact that Latour would object to observing the
actor network as a mathematical network \cite{key-11}.) Is the dead
body of Laura Palmer a character? Is the photo of Laura Palmer, and
is it the same one? How about the Roadhouse?

\textcolor{black}{Furthermore, this study largely ignored the multiplicity
of common appearances of characters in scenes. For instance, in our
count, Truman and Cooper share almost 50 common appearances, often
in scenes that involve other characters as well. This makes them heavily
correlated and they may appear as a single entity, a strongly connected
pair. Studying correlations of this sort is a natural extension of
the work presented here.}

The one takeaway that stands out and presents a statistically robust
result is The Dale Cooper Effect. The sharp demarcation between the
two networks embodied by the protagonist introduced as a median character
is the peculiar shape of story. Do similar structures emerge in other
works of fiction? What is the fundamental relationship of The Dale
Cooper Effect with storytelling? Those questions guide our future
work.

\section*{Acknowledgements}

The idea of this paper came from a talk I prepared for the conference
\emph{Beyond Life and Death: Twin Peaks at Thirty.} Discussions with
Jackie Brown, Anastasia Tsukanova, Tom O'Dea, and Dennis McNulty made
the paper much better than it originally was. \textcolor{black}{Support
from Science Foundation Ireland (SFI), co-funded under the European
Regional Development Fund under Grant Number 13/RC/2077\textbackslash\_P2
is acknowledged.}

\end{document}